\documentclass[]{aa}   
\usepackage{hyperref}
\usepackage{graphicx}
\usepackage{txfonts}
\usepackage{ulem}
\begin{document}
\title{Detection of quasi-periodic oscillations in the 37 GHz radio light curve of the blazar Ton 599 during 1990 -- 2020}
\titlerunning{Radio QPO in the Blazar Ton 599}
   \author{
        Alok C.\ Gupta \inst{1,2} 
        \thanks{\tt Corresponding authors: acgupta30@gmail.com (ACG), cuilang@xao.ac.cn (LC)}
          \and
          Alexandr E. Volvach\inst{3}
          \and
        Shubham Kishore\inst{2,4}
          \and
          Larisa N. Volvach\inst{3}
          \and
        Paul J.\ Wiita\inst{5}
        \and
        Lang Cui\inst{1} 
        \textsuperscript{$\star$} 
          \and \\
        Mauri J. Valtonen\inst{6,7}
          \and
        Sandeep K. Mondal\inst{8}
          \and
        Haritma Gaur\inst{2}
    }
    \authorrunning{Gupta et al.}
   \institute{
        Xinjiang Astronomical Observatory, Chinese Academy of Sciences, 150 Science-1 Street, Urumqi 830011, People's Republic of China
        \and
   Aryabhatta Research Institute of Observational Sciences (ARIES), Manora Peak, Nainital 263001, India
   \and
   Radio Astronomy Laboratory of Crimean Astrophysical Observatory, Katsively RT-22, Crimea
   \and
   Indian Institute of Astrophysics (IIA), 2nd Block, Koramangala, Bangalore 560034, India
   \and
    Department of Physics, The College of New Jersey, 2000 Pennington Rd., Ewing, NJ 08628-0718, USA 
    \and
    FINCA, University of Turku, FI-20014 Turku, Finland
    \and
    Tuorla Observatory, Department of Physics and Astronomy, University of Turku, FI-20014 Turku, Finland
    \and
    Tsung-Dao Lee Institute, Shanghai Jiao Tong University, 1 Lisuo Road, Shanghai, 201210, People’s Republic of China
        }
         
  \abstract
   {Blazars are a subclass of radio-loud active galactic nuclei (AGN) that display strong multi-wavelength variability on diverse timescales ranging from years down to minutes. In the last 1.5 decades, there have been occasional detections of quasi-periodic oscillations in several blazars in their time series data.}
   {We searched for quasi-periodic oscillations (QPOs) in the 37 GHz radio band light curve of the flat-spectrum radio quasar Ton~599 made at the RT-22 radio telescope in Simeiz, Crimea, from 1990 to 2020. We also searched for QPOs in the available gamma-ray and optical data during the time span of these radio observations.}
   {To identify and quantify the QPO nature of this radio light curve of Ton 599, we used the Lomb–Scargle periodogram (LSP), REDFIT, and weighted wavelet Z-transform (WWZ) analyses. We performed LSP analyses of the gamma-ray and optical data.}
   {We report the detection of a likely QPO of about 2.4 years in a portion of the 37 GHz radio light curve of Ton 599. No QPO signatures of similar timescales were found in either the $\gamma$-ray or optical (R-band) wavebands.}
   {We briefly discuss possible emission models for radio-loud AGN that could explain such QPOs with periods of a few years.}
    

\maketitle
 
\section{Introduction}
\label{section1}
\noindent
Blazars are a subclass of radio-loud (RL) active galactic nuclei (AGN) that emit radiation across the entire electromagnetic (EM) spectrum from radio to gamma-ray bands. Blazars possess relativistic charged particle jets pointed almost in the direction ($\leq \rm{10}^{\circ}$) of the line of sight (LOS) of the observer \citep{1995PASP..107..803U}. Because of the very small viewing angle, relativistic effects result in substantially magnified observed emissions, such that the jet emission dominates the overall observed fluxes from blazars in nearly all bands. Throughout the whole EM spectrum, blazars show remarkable flux, spectral, and polarization variability, with emission dominated by non-thermal processes \citep[e.g.][and references therein]{1993ApJ...411..614U, 2010ApJ...716...30A, 2015ApJ...807...79H, 2017Natur.552..374R, 2023ApJ...957L..11G}. Blazars are the collective term for flat spectrum radio quasars (FSRQs) and BL Lacerate objects (BLLs). While FSRQs exhibit strong emission lines in the composite optical/UV  spectrum \citep{1978PhyS...17..265B, 1997A&A...327...61G}, BLLs display featureless or extremely faint emission lines with equivalent widths (EWs) $\leq$ 5\AA \citep{1991ApJS...76..813S, 1996MNRAS.281..425M}. Blazar spectral energy distributions (SEDs) display a double hump structure in which the low-energy and high-energy humps respectively peak in infrared (IR) to X-ray bands and gamma-ray energies \citep{1998MNRAS.299..433F}. The low-energy and high-energy parts of the SEDs originate from relativistic electrons in the jet, and are commonly explained by synchrotron radiation and inverse Compton (IC) emission, respectively \citep[e.g.][and references therein]{2007Ap&SS.309...95B, 2017MNRAS.472..788G}. \\
\\
Periodic oscillations, or quasi-periodic oscillations (QPOs), have been observed quite often in the light curves (LCs) of neutron stars and stellar-mass black hole (BH) binaries \citep{2006ARA&A..44...49R}. While the LCs of AGN are mostly non-periodic across the entire EM spectrum, detections of periodic oscillations or QPOs in those of blazars and other subclasses of AGN with quite different periods have been claimed in the literature \citep[e.g.][and references therein]{1985Natur.314..148V, 1989ApJ...347..171F, 1993Natur.361..233P, 1998MNRAS.295L..20I}. More recently, the search for QPOs in different subclasses of AGN LCs has been performed using various new analysis techniques after good evidence of a QPO of period $\sim$1 hour was found in the X-ray LC of the AGN RE J1034+396 \citep{2008Natur.455..369G}. Some blazars have been found to have occasional QPO detections in a variety of EM bands and with quite a wide range of periods \citep[e.g.][and references therein]{2009ApJ...690..216G,2013MNRAS.436L.114K,2015ApJ...813L..41A,2021MNRAS.501...50S,2022Natur.609..265J, 2022ApJ...926L..35O,2022MNRAS.510.3641R,2023ApJ...950..173D,2024ApJ...977..166T}.
QPOs have also been claimed in other AGN subtypes \citep[e.g.][and references therein]{2008Natur.455..369G,2014MNRAS.445L..16A,2015MNRAS.449..467A,2015Natur.518...74G,2016ApJ...819L..19P,2018A&A...616L...6G}. \\ 
\\
Ton 599 is a luminous FSRQ and is also known as  4C~29.45, and as 1156+295 ($\alpha_{2000.0} = \rm{11}^{h} \ \rm{59}^{m} \ \rm{32.07}^{s}, \ \delta_{2000.0} = +\rm{29}^{\circ} \ \rm{14}^{'} \ \rm{42.0}^{"}$),\footnote{\url{https://www.lsw.uni-heidelberg.de/projects/extragalactic/charts/1156+295.html}} located at redshift, $z = 0.72469$ \citep{2010MNRAS.405.2302H}. It shows significant flux, polarization, and spectral variability throughout the EM spectrum from radio to $\gamma$-rays \citep[e.g.][]{2004A&A...417..887H,2013A&A...555A.134L,2014ApJ...791...53A,2018ApJ...866..102P,2019ApJ...871..101P,2022ApJ...926..180H,2024MNRAS.529.1356M}. A huge multi-wavelength (MW) outburst in late 2017 led to its first discovery at very high-energy (VHE) $\gamma$-rays by the Very Energetic Radiation Imaging Telescope Array System (VERITAS) and Major Atmospheric Gamma Imaging Cherenkov (MAGIC) telescopes \citep{2017ATel11061....1M,2017ATel11075....1M,2026MNRAS.tmp...66A}. Later studies of this outburst were focused on restricting the jet’s physical properties and the location of the emission zone, revealing a complex link between the non-thermal continuum and the broad emission lines \citep{2018ApJ...866..102P,2019ApJ...871..101P,2022ApJ...926..180H}. Highly superluminal motion has been revealed by VLBI studies of Ton~599's parsec-scale jet, enabling estimates of the jet viewing angle $\Theta \leq \rm{2.5}^{\circ}$ and Doppler factor $\delta = 12 \pm 3$ \citep{2017ApJ...846...98J}.  Using the Mg~II emission line, the mass of the supermassive black hole (SMBH) at the centre of this FSRQ is estimated to be $9 \times 10^{8} \rm{M}_{\odot}$ \citep{2019PhDT.......127K}. \\
\\
Hints of a QPO with a period of 3.29 years were detected in the 22 GHz radio LC of this source between 1984 and 2004 using a Lomb--Scargle periodogram (LSP), which was in rough agreement with the 3.49 yr feature seen in the discrete correlation function;  the structure function gave a variability timescale of 1.21 yr \citep{2007A&A...469..899H}.  The data for that study were taken with the 14-meter Mets{\"a}hovi Radio Telescope in Finland. Multiple QPOs, with characteristic periods of $\sim$1.7 (P4), $\sim$2.4 (P3), $\sim$3.4 (P2), and $\sim$7.5 years (P1) were claimed to be detected in 4.8, 8.0, and 14.5 GHz quasi-simultaneous radio observations of the source from 1980 to 2010 from the University of Michigan Radio Observatory (UMRAO), USA \citep{2014MNRAS.443...58W}. These observations suggest a harmonic relationship in frequency of 4:3:2:1, with f1 as the fundamental frequency \citep{2014MNRAS.443...58W}. \\
\\
In this paper we report the detection of a QPO with period of $\sim$2.4 years in the 37 GHz radio LC of the blazar Ton 599 taken from 1990 to 2020. QPOs with timescales of months to years are believed to be an important tool for predicting the next outburst time in a blazar in a specific waveband of the EM spectrum. We discuss possible explanations for this finding, including AGN models that involve a binary  SMBH system \citep[e.g.][and references therein] {1996ApJ...460..207L,1996A&A...305L..17S,2024ApJ...968L..17V},  a helical structure in the jet \citep{2004ApJ...615L...5R,2015ApJ...805...91M}, or Lens–Thirring precession \citep{1998ApJ...492L..59S,2018MNRAS.474L..81L}. \\
\\
In Section 2 we provide information about our 37 GHz radio observational data and its analysis. Because these radio results are our prime focus, and as we found null results in the $\gamma$-ray and optical bands, the details about the data acquisition and analysis for those bands are provided in  Appendix \ref{optical_gamma}. In Section 3 we present the results of the radio observations, and a discussion and conclusions are presented in Section 4. Details about the analysis techniques are found in the Appendices.
 
\begin{figure*}
    \centering
    \includegraphics[width=1.0\linewidth, angle=0, trim= 0 .05cm 0 0, clip]{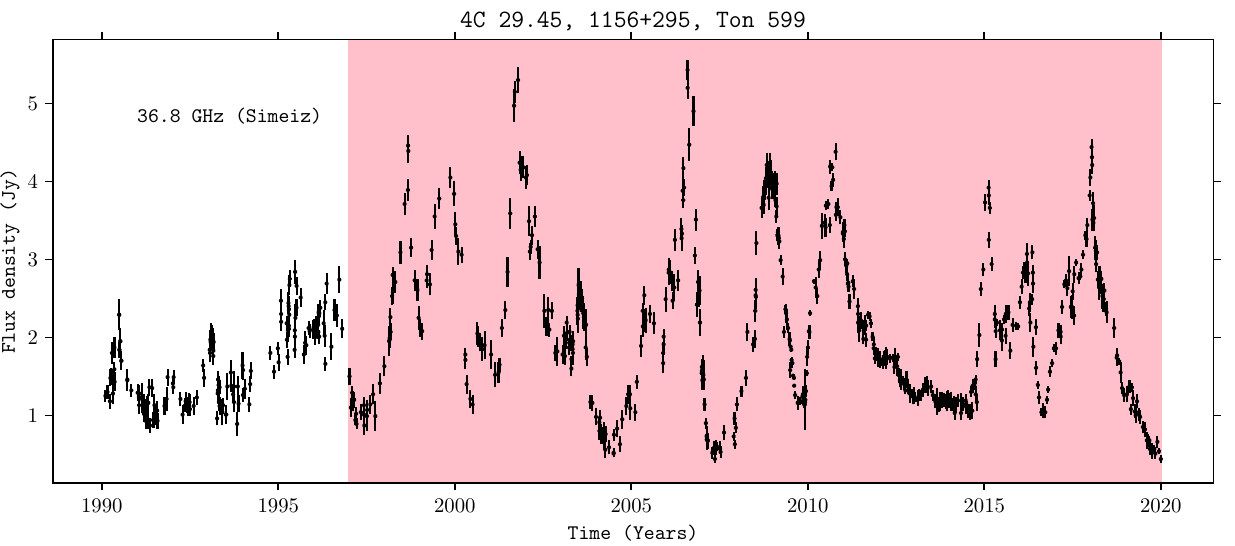}
    \caption{Light curve of Ton~599 (yr: 1990 -- 2020), highlighting the portion of interest for periodicity searches.}
    \label{fig:1}
\end{figure*}
 
\section{Radio observations and analysis}
\label{section2}
\noindent 
Observations on the RT-22 radio telescope in Simeiz were carried out at a frequency of 36.8 GHz using a modulation radiometer in the beam modulation mode \citep{2024A&A...691L...9V}. The width of the main lobe of the radio telescope beam pattern was 100 arc seconds. The pointing and tracking accuracy of the instrument, taking into account systematic errors, was several arc seconds. \\
\\
Before measuring the radiation flux density, the position of the source was refined by scanning in right ascension and declination. Then the radio telescope was tuned to the source using both input horns with mutually orthogonal polarizations in turn. The antenna temperature of the source response was determined as the difference between the radiometer responses averaged over 30 seconds in the two specified antenna positions. Depending on the signal-to-noise ratio of the radio source response, a series of 30–60 measurements was carried out, after which the average signal value was calculated and its root mean square error was estimated. The orthogonal polarization of the lobes allowed the total radiation intensity from the source to be measured regardless of its polarization.\\
\\
The absorption of radiation in the Earth's atmosphere was taken into account using the atmosphere cuts method, recording the differences in antenna temperatures at given elevation angles.
The measured values of antenna temperatures, corrected for atmospheric absorption, were recalculated into spectral flux density based on observations of calibration sources such as DR21, 3C 274, Jupiter, and Saturn. The conversion of antenna temperatures into spectral flux density was performed taking into account the dependence of the effective antenna area \( A_{\text{eff}} \) on the elevation angle \( h \). \\
\\
Figure\ \ref{fig:1} displays the 36.8\ GHz LC of TON\ 599 from 1990 to 2020, depicting notable oscillatory features from 1997 onwards. To search for any periodicity, we first performed the weighted wavelet z-transform (WWZ) analysis to get a 2D overview of the LC nature comprising the temporal and frequency domains. The existence of the peculiarity seen was further tested by two other techniques, which employ distinct approaches, used traditionally: fitting the generalized Lomb--Scargle periodogram (LSP), and fitting the time series with the first-order autoregressive (AR(1)) method, using REDFIT software. The LSP method helps to assess the power spectra of the time series by fitting multiple sinusoidal functions, and is able to detect any inherent periodicity and test its significance. The AR(n) approach utilizes the hypothesis that a particular data value in the time series is a manifestation of previous $n$ data values (here for AR(1), $n=1$), estimating a set of coefficients to compute a theoretical power spectrum of the time series. This spectrum is then used to get separate significance levels for {any putative QPOs. Details of the WWZ, LSP, and AR(1) methods are} explained in the Appendices.
  
\section{Results}
\label{section3}
We found a strong periodic signal of around 2.4 years, consistent in all the employed analysis approaches for the post-1997 portion of the LC. The WWZ analysis reveals that the signal detection of this period is almost unaffected by edge effects due to the finiteness of the LCs (see Fig.~\ref{fig:2} and the corresponding section of the Appendix); there are at least  six cycles. Looking at the 2D WWZ map, the nature of this prominence can be considered to be both transient and quasi-periodic. Over a visually constricted portion of the LC from yr~1997 to 2020, the LSP and the REDFIT methods both agree with the frequency detection of $\sim$0.42~yr$^{-1}$. The significance of the detection has been achieved with both the LSP and AR(1) techniques, and each with two distinct methods: one following the $\chi^2$ distribution approach, and the other with the percentile of the simulated LCs' power spectra (see the corresponding parts of the Appendix). Within the span between 1997 to 2020 this frequency peak has approximately 99.9\% significance using the LSP, and over 99.73\% for the AR(1) analysis. We also computed the global significance of the peak, taking into account the number of independent frequency trials, and found it to be around 95\%. \\
\\ 
We also analysed the entire LC and an even more limited portion, following \cite{2022ApJ...926L..35O}.  For the entire LC the local significance of the peak using the  LSP analysis are around 99.773\% ($\chi^2$) and 99.868\% (percentile of simulation); globally, they are around 91.31\% and 94.85\%, respectively. If we only consider the more limited portion of the LC between 2006 and 2020 the local significance levels of this peak are 99.9976\% ($\chi^2$) and 99.9984\% (percentile of simulation); globally they are 99.936\% and 99.904\%, respectively.\\
\\ 
We also examined the published data taken during  much of the same interval in the $\gamma$-ray and optical bands, but found no obvious QPO period of similar timescales in either of those bands. The details of the data analysis and the corresponding results are presented in Appendix \ref{optical_gamma}.

\section{Discussion and conclusions}
\label{section5}
\noindent
In this paper we used the 37 GHz radio band LC of the FSRQ Ton 599 to look for any QPOs over a 30 year time span (1990--2020). These observations were all made at the RT-22 radio telescope at Simeiz, Crimea. To search for QPOs, we employed weighted wavelet Z-transform (WWZ), generalized Lomb-Scargle periodogram (GLSP), and  REDFIT analyses.  Each of these methods yielded strong indications of a QPO with a period of $\sim$2.4 years.  Together, at least nine QPO cycles are visible to the naked eye as distinct peaks in Fig.~\ref{fig:1}. The 2.4 yr QPO is found to cross the nominal 99.73\% significance levels, regardless of the approach employed in GLSP and REDFIT, and has a global significance of $>$99.9\% in the more limited span between 2006 and 2020. In addition to these features, there are some higher-frequency peaks in the LSP (Fig.\ \ref{fig:3} and Fig.\ \ref{fig:4}) that could be seen to cross the corresponding significance levels. However, they mainly lie in the region where instrumental white noise dominates, hence are more likely due to instrumental artefacts. In addition, these features have very low periodogram powers, and the WWZ plot (Fig.\ \ref{fig:2}) shows no substantial contributions from these features.\\ 
\\
In addition to the radio QPO analysis, we also considered long-term observations of Ton 599 in the optical R-band and $\gamma$-ray (0.1--300~GeV) to check for the presence of QPO features with periodicities similar to that which appears to be present in the radio band. The data we examined in these bands are briefly discussed in  Appendix~\ref{optical_gamma}. Though neither of these light curves fully covers the 1990--2020 period that would allow  a complete comparison, they span substantial portions of it. We followed \cite{2025A&A...703A.259V} to examine extensive R-band data covering 2011--2020. Their analysis showed an essentially power-law PSD with no indication of a QPO, and our reanalysis yielded similar results.  The $\gamma$-ray light curve was obtained from the Fermi Large Area Telescope (Fermi-LAT) \citep{2009ApJ...697.1071A} for the interval spanning 2008--2020. Over their entire temporal coverages both of these light curves exhibit strong variability. However, they did not show any peculiar QPO features anywhere around the period of 2.4 years.  This should not be surprising since in most blazar models the radio band emission does not arise from the same regions as the optical and $\gamma$-ray fluxes do. Models involving IC processes typically upscatter optical photons into the $\gamma$-ray band, so correlations between those two bands are more common than those with the radio bands \citep[see reviews by][]{2019ARA&A..57..467B,2025A&ARv..33....8R}.  Some blazars do exhibit correlated variations between the  $\gamma$-ray and millimetre-radio bands, with the radio typically lagging the $\gamma$-ray by months, indicating that in those cases these emissions emerge from nearby portions of the jet \citep[e.g.][]{2011A&A...532A.146L}.  As we do not see such an interband correlation in Ton 599, the simplest explanation is that the $\gamma$-ray and radio emission come from different parts of the relativistic jet. \\
\\
There have been several explicit attempts to search for simultaneous multi-wavelength QPOs in blazars. Some evidence of simultaneous $\gamma$-ray and optical QPOs with a period of a few tens of days were found for a few objects \citep[e.g.][and references therein]{2016ApJ...820...20S,2017A&A...600A.132S,2020A&A...642A.129S,2021MNRAS.501...50S}. Quite strong evidence of simultaneous QPOs in optical flux (and linear polarization) and $\gamma$-ray flux, with cycles as short as approximately 13 hours, has been reported in BL Lacertae during the highest portion of an outburst \citep{2022Natur.609..265J}. These QPO properties could be explained by current-driven kink instabilities \citep{2020MNRAS.494.1817D}. \citet{2017ApJ...847....7B} searched for 15 GHz radio and gamma-ray QPOs in the blazar PKS 0219$-$164 but only found a QPO in the radio band with a period of 270$\pm$26 days, along with its two harmonics; they discussed various AGN emission models that might explain that result. \citet{2015ApJ...813L..41A} searched for a multi-wavelength QPO in the blazar PG 1553+113; however, their X-ray data was too sparse and poorly sampled, and their radio data only covered part of the duration of their $\gamma$-ray and optical data. Strong evidence of a simultaneous QPO in $\gamma$-ray and optical bands was found, while the oscillation in the radio band was less regular than that seen in the other bands. \\ 
\\
The cumulative emission of blazars in all EM bands is dominated by the non-thermal emission from jets and quasi-thermal emission from accretion disks. Both fluctuations of accretion disk origin advected into the jet and those with solely jet-based origins could produce the $\sim$2.4 yr QPO found here in the 37 GHz radio LC of the FSRQ Ton 599. The establishment and expansion of the instabilities in the jets could be impacted by time-dependent variations in the fuelling of the jet by the central engine (SMBH and accretion disk). The jet, which is highly Doppler boosted due to its low inclination angle \citep[$\sim$2.5$^{\circ}$;][]{2017ApJ...846...98J}, is unquestionably the source of the emission at the 37 GHz radio frequency used in this study. Consequently, internal jet activities are far more likely to be the cause of any QPO seen than accretion disk processes.\\
\\
In the 1984--2004 22 GHz radio LC  of Ton 599 \citep{2007A&A...469..899H}, a visual inspection does not show any QPOs until 1995. This is reflected in their LSP in which we note that merely three cycles of a period of $\sim$3.29 years might be present. The differences in our results with those of \citet{2007A&A...469..899H} are not surprising in that there is no real overlap between the timespan of the QPO searches in these two studies. \citet{2007A&A...469..899H} use the data from 1984 to 2005 for their QPO search, but we found the most significant QPO signal, with a period of $\sim$2.4 years, in the data from 2006 to 2020.
In the 1980--2010 radio LCs at 4.8, 8.0, and 14.5 GHz, once again the source does not show any QPOs until 1995. However, multiple periods were possibly seen in a LSP analysis of these 1995--2010 radio light curves, but only about five cycles of the strongest frequency were present \citep{2014MNRAS.443...58W}. \citet{2025A&A...693A.318K} recently presented the 37 GHz LC of Ton 599 taken between 1985 and 2025 at the Mets{\"a}hovi Radio Observatory, Finland. Their PSD analysis yielded a bending power law with a bend at about 1500 days, and the best-fit slope $\beta =$ 2.5 indicates relatively slow variability. The radio data of \citet{2025A&A...693A.318K} do overlap with ours, but they have not performed a dedicated search for QPO signatures. They performed an overall periodogram study, and it is worth noting that a mixture of aperiodic, periodic, and stochastic variations in the light curve can greatly reduce the peak significance of any genuine periodic variation, which could have resulted in their not noting any QPOs. Since \citet{2025A&A...693A.318K}  also made their observations at 37 GHz, we note that the same type of variations we  are discussing  here can be seen in the segment of their light curve (their Fig.\ A.9) between 2006 and 2020, which exhibited the strongest QPO signature in the present study.\\
\\
One of the most plausible scenarios for the presence of QPOs on multi-year timescales in blazars is the SMBH binary model, which is well established for the blazar OJ 287 through the detection of its predicted outbursts, separated by $\sim$12 years,  since 1996 \citep[e.g.][]{1988ApJ...325..628S,1996A&A...305L..17S,1996ApJ...460..207L,1997ApJ...484..180S,2006ApJ...643L...9V,2006ApJ...646...36V,2024ApJ...968L..17V,2023ApJ...957L..11G}. In this model the variable radio flux is related to differences in the accretion rate and injection into the jets, which are induced by the presence of the secondary black hole. Due to the precession of the binary orbit, these variations are not strictly periodic, only quasi-periodic. In addition, there is a drift of the quasi-periodic component over several cycles \citep{2006ApJ...646...36V}. The amplitude  also varies  \citep{1997ApJ...484..180S}. The latter may be used to find the precessional period of the major axis of the binary orbit. This mechanism could, in principle, also explain the case of Ton 599; however, an observed $\sim$2.4 yr period is certainly shorter than expected for a SMBHB orbit, and the change in QPO period is faster in Ton 599 than in OJ 287.\\
\\
Thus, our results  better support the general class of models originating in geometric features of the approaching jet. Such models include a curved jet \citep[e.g.][]{1999A&A...347...30V}, a twisted inhomogeneous jet \citep[e.g.][]{2017Natur.552..374R}, and a wiggling jet \citep[e.g.][]{2024A&A...692A..48R}.  These mechanisms could explain the long-term radio activity of the blazar Ton 599, including a transient QPO. Either Kelvin--Helmholtz or current-driven kink instabilities \citep{2020MNRAS.494.1817D} in the plasma can create an inhomogeneous jet or filament. Those instabilities create a rotating, curving helix-like jet structure that emits radiation from different regions along the jet, variably amplified by changing Doppler beaming. High-resolution observations, similar to the ones shown for the FSRQ blazar 3C 279 \citep{2023NatAs...7.1359F}, can resolve the radio emissions as we get farther away from the downstream point where they are generated. Relativistic frame-dragging may also be responsible for inducing wiggling of the jet through the Lens–Thirring precession in the inner region of the accretion disk, from which the jet may be launched \citep{1998ApJ...492L..59S,2018MNRAS.474L..81L}. These features of inhomogeneous or wiggling jets can explain QPOs detected from up to several months to a few years, such as the signals seen in Ton 599 at a period of 2.4 years. \\ 
\\
Other geometrical origins or variations of those mentioned above  might also produce QPOs with periods of a few years.  These scenarios include helical trajectories in extragalactic radio jets caused by differential Doppler boosting effects \citep{2004ApJ...615L...5R}. That paper shows that light-travel time effects would cause the observed period for radial helical motion to be significantly less than the actual physical driving period, but periods greater than 1 year may be linked to periodicity caused by Newtonian-driven precession \citep{2004ApJ...615L...5R}. Another possible explanation involves special relativistic jets with an emitting blob in a conical geometry \citep{1992A&A...255...59C,2015ApJ...805...91M}. A related and very interesting possibility is a fully relativistic model in which orbiting blobs are in helical motion along a funnel or cone-shaped magnetic surface anchored to the accretion disk near the SMBH. It is capable of producing QPOs of the order of years \citep{2015ApJ...805...91M,2020Galax...8...67R}.

\begin{acknowledgements}
\noindent
We thank the reviewer for useful comments that helped us to improve the manuscript. This work was supported by the CAS `Light of West China' Program with Grant No.~2021-XBQNXZ-005. ACG is partially supported by the CAS `President's International Fellowship Initiative (PIFI)' with Grant No.~2026PVA0040. LC acknowledges the support from the Tianshan Talent Training Program with Grant No.\ 2023TSYCCX0099. HG acknowledges the financial support from Science and Engineering Research (SERB), India, through SERB Research Scientist award SB/SRS/2022-23/113/PS at ARIES, Nainital, India.
\end{acknowledgements} 
\bibliographystyle{aa}
\bibliography{ref}
 \begin{appendix}\label{apx}
\section{Weighted wavelet Z (WWZ) analysis}
\begin{figure}[h]
    \centering    \includegraphics[width=1.\linewidth]{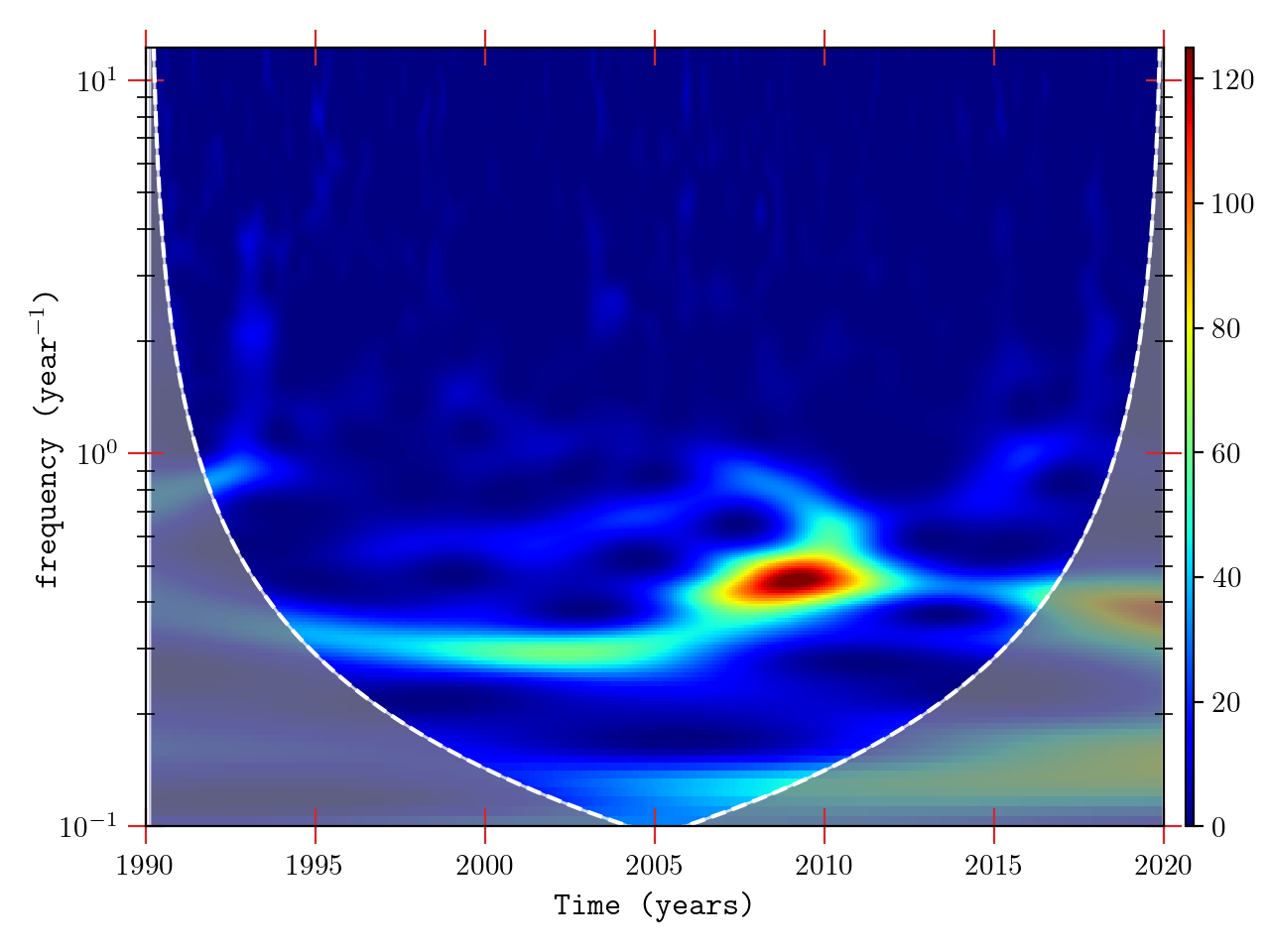}
    \caption{WWZ map of light curve of Ton~599. The low-resolution region below the dashed white line represents the ROI.}
    \label{fig:2}
\end{figure}
\noindent
Visual inspection of the LC in Fig.~\ref{fig:1} reveals notable oscillatory features beginning at around the year~1995 and persistent thereafter. We used the weighted wavelet z-transform (WWZ) analysis, a robust technique used to assess both the temporal and frequency domains of the LC. In this analysis, periodicity searches are made with time-localized continuous functions (wavelets), defined by two parameters, scale and location. Depending on the wavelets used, the scales are related to the oscillatory time periods in the light curves \citep[e.g.][]{1996AJ....112.1709F}. We used the Morlet wavelet (given as $f(z)=e^{-cz^2}(e^{iz}-e^{-1/4c})$) as the mother wavelet. The WWZ technique involves the product of a sine wave with a Gaussian envelope and decomposes a LC into a two-dimensional mesh of time and frequency components. Taking $z=\omega(t-\tau)$, the WWZ map is given as 
\begin{equation}
    W[\omega, \tau : x(t_\alpha)] = \omega^{1/2} \sum_{\alpha=1}^{N}x(t_\alpha)f^*[\omega(t_\alpha-\tau)],\\
    \label{DWT_form}
\end{equation}
where $\omega$, $\tau$, $x(t_\alpha)$, and $N$ denote the angular frequency, location of the wavelet, flux at the timestamp $t_\alpha$, and the total number of data points, respectively.\\
\\
Figure~\ref{fig:2} shows the 2D WWZ map of the LC in Fig.~\ref{fig:1}. An edge effect, which is typical of time series data analyses, also comes into play when dealing with the Morlet wavelet, and it yields a region of influence (ROI) \citep[see][for details]{1998BAMS...79...61T}. The shaded region under the white dashed lines in Fig.~\ref{fig:2} represents the ROI, the region in the 2D map where the edge effects are dominant due to the finite length of the light curve.\\
\\
In our analysis, we have seen that the strongest signal is at around a frequency of 0.42 yr\(^{-1}\). The signal is well outside the ROI, lasting for around 6 cycles within the ROI, and thus is probably a reliable QPO. Considering the shape of the feature in the WWZ-plot, this feature can be characterized as transient and quasi-periodic. It can also be noticed that the WWZ map traces this feature as a progression from a feature with a longer period (shorter frequency value) to one with a shorter QPO period. One of the nice features of this 2D map is that the estimation of an average power spectrum, or time-averaged periodogram (TAP), between any two epochs of the observations, is a byproduct. The TAP curves estimated for the  span between 1997 and 2020 and also for just 2006 -- 2020 are presented in Fig.~\ref{fig:3}. We see that  the most elevated values of these TAPs are at \(\sim\)0.42 yr\(^{-1}\) agreeing with those of the LSPs (see Fig.~\ref{fig:3}).
 
\section{Generalized Lomb-Scargle periodogram (LSP) analysis}
\begin{figure}
    \centering
    \includegraphics[width=1\linewidth]{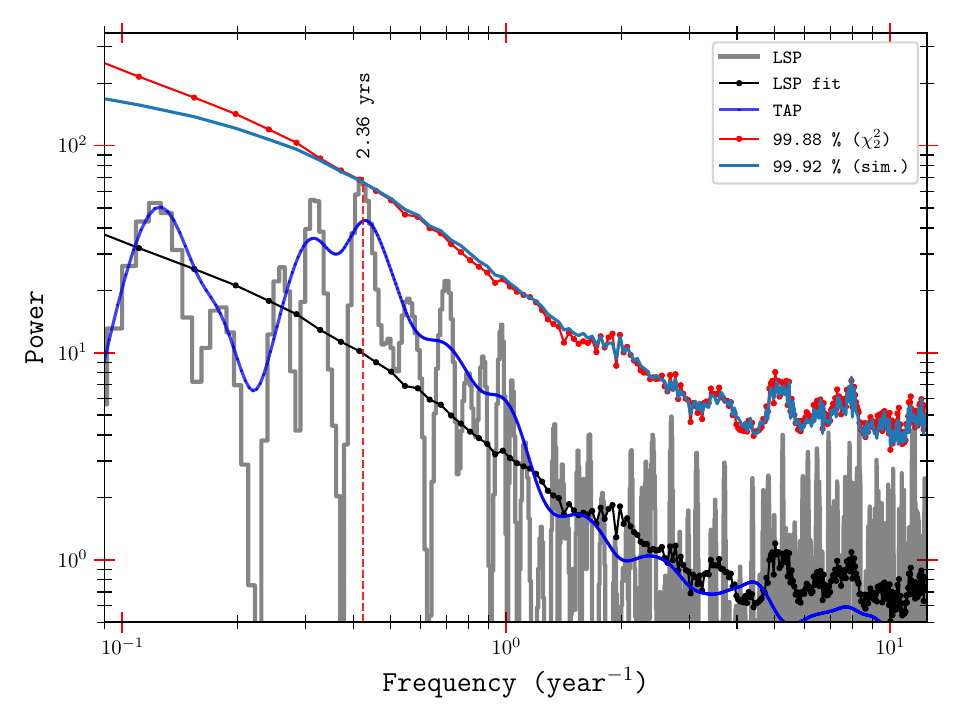}\\
    \includegraphics[width=1.0\linewidth]{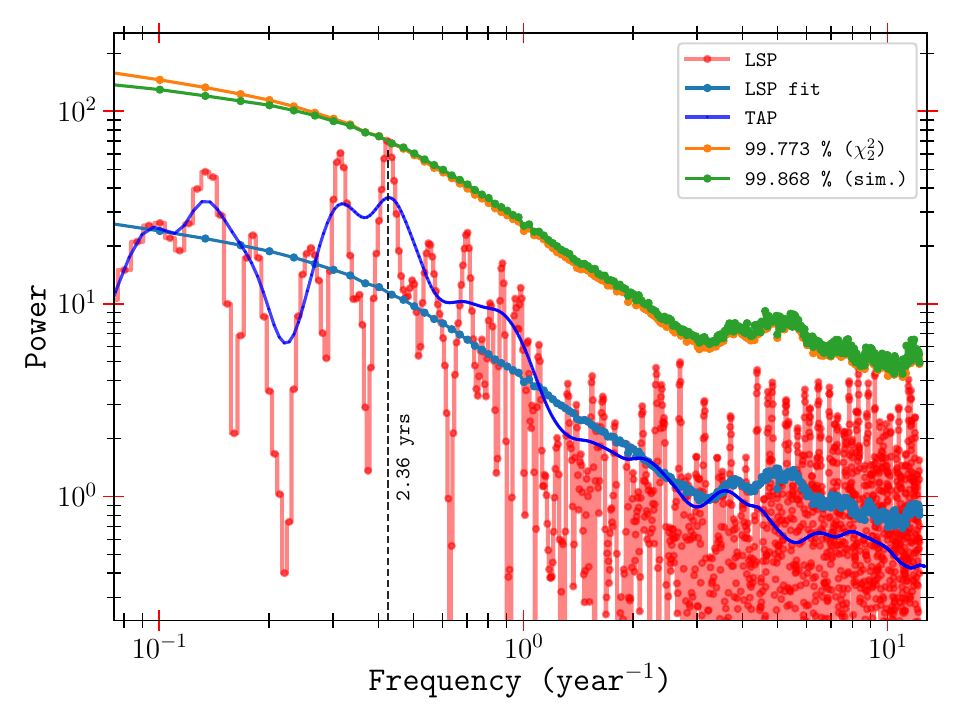}\\
    \includegraphics[width=1.075\linewidth,trim=0 0 0 1cm,clip]{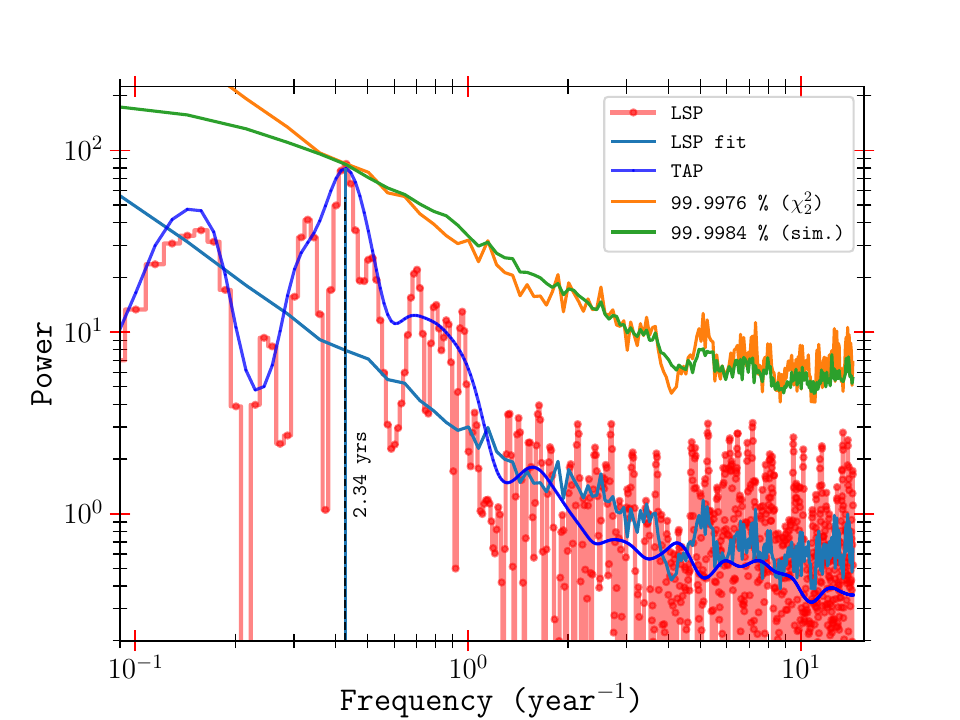}
    \caption{Upper panel: LSP of the main segment of the light curve of Ton~599 from 1997 to 2020. The LSP fit is the mean of LSPs from $10^6$ simulated light curves; the 99.88\% \(\chi^2\) significance curve is evaluated following \citet{2005A&A...431..391V} (see corresponding Eq.\ 16); the 99.92\% (sim.) shows the percentile significance evaluated from the distribution of the simulated light curves' LSP powers. The QPO signal has been found to be at around the noted significance levels. Middle panel: As above, but for the entire set of observations from 1990 to 2020.  Lower panel: As above, but for the restricted span 2006 to 2020.}
    \label{fig:3}
\end{figure}
\noindent
The power spectral density (PSD) analysis offers an approach to deduce information about the source of the variability, as different physical models produce different ranges of spectral slopes \citep[][]{2016ApJ...820...12P, 2019ApJ...877..151W}. In addition, any signature of periodicity or quasi-periodicity appears as a distinct peak in the PSD. The blazar light curves are of a stochastic nature; hence, it is customary to test the periodicity results with techniques other than the WWZ analysis. One such method is the generalized periodogram (GLSP) \citep[e.g.][]{2018ApJS..236...16V}, a common tool in the frequency analysis of time series data that estimates the power at given sinusoidal frequency values and is equivalent to least-squares fitting of sine waves. With the help of the WWZ map, we focus on the more obviously oscillatory portion of the LC from 1997 to 2020 for the LSP analysis.\\ 
\\ 
We have followed the publicly available Python package {\tt PyAstronomy} \citep{2019ascl.soft06010C}\footnote{https://github.com/sczesla/PyAstronomy} to compute the periodogram of the LC. Figure~\ref{fig:3} includes the periodogram resulting from the GLSP analysis that shows major power enhancements at the  frequency 0.42~yr\(^{-1}\) emerging from the WWZ analysis. We tested the significance  using the LC simulations based on the method described in  \cite{2013MNRAS.433..907E}, where  the power spectral density (PSD) of the LC was fitted with the bending power law model given as 
\begin{equation}
    P(\nu)=A\frac{\nu^{-a_l}}{1+(\nu/\nu_b)^{a_h-a_l}}+c \ ,
    \label{bending power law}
\end{equation}
where \(A,~\nu_b,~a_l,~a_h,~{\text{and}}~c\), respectively, are the normalization, bending frequency, lower and upper frequency index, and offset. The method utilizes the PSD and flux distribution of the input LC  to simulate artificial LCs with similar statistical properties.  With the fitting parameters of the estimated PSD and the flux distribution,  we simulated a total of 10$^{6}$ artificial LCs and calculated the generalized periodogram of each individual light curve. This way, we have a distribution of 10$^{6}$ data points pertaining to each individual periodogram frequency. The mean of the 10$^{6}$ GLSPs, considered to be the underlying spectrum (LSP fit), is depicted by the black curve in Fig.~\ref{fig:3}.
The TAP from the WWZ analysis is also plotted in Fig.\ \ref{fig:3} to show the agreement of frequencies where the powers have a local maximum. \\ 
\\
In order to estimate the significance level of a peculiar periodic signal, we followed two approaches. The first one includes the estimation of levels by finding the number of times the observed LSP peak is greater than the LSP powers of all simulated light curves. With this method, the 2.36 yr QPO has been found to have a significance of around  $99.92\%$. The other method includes finding proper multiplication factors to the mean LSP for different probability levels \citep[the $\chi^2$ approach; see e.g.][]{2005A&A...431..391V}. These are helpful to get nominal local significance levels, but one must also consider the number of independent trials to get a global significance level.  We used Eq.\ (16) of \citet{2005A&A...431..391V}, given as \(\gamma_\epsilon=-2\ ln[1-(1-\epsilon)^{1/n}]\), to go from local to global significance estimations. Here $n$ is the number of independent trials and $\epsilon$ is the probability of getting a significant peak under the null hypothesis. The different values of $\gamma_\epsilon$ offer estimates for different probability levels for $n$ independent frequency examinations. The ratio $k=\gamma_\epsilon(n)/\gamma_\epsilon(n=1)$ acts as a simple means to access global levels using the estimated local significance levels. Visual inspection of Fig.\ \ref{fig:3} reveals that the light curve is dominated by instrumental white noise for fluctuations beyond $\geq2$ yr$^{-1}$ with a transition phase around 1--2 yr$^{-1}$; searches in that region could produce spurious periodicity detections, so the number of independent frequency trials was limited to below 2 yr$^{-1}$ ($n \approx30-45$). Within this range, we find that the 2.36 yr QPO, which has significance of 99.88\% locally, has a global significance level in the range 94.7\% -- 96.5\%.\\
\\
The normalized power of a real (periodic or quasi-periodic) signal in the PSD is sensitive to aperiodic components or components with differing periods in the LCs, and is naturally diminished by their existence. This can significantly alter the actual significance level of the peak. Thus it can be reasonable to consider further segmentation of the LC, including all of the LC and also focusing on the epochs in which the signal appears stronger (estimated via the WWZ map) \citep[see e.g.][]{2022ApJ...926L..35O}. When the entire LC is analysed the signal is weaker, peaking around 2.3 yr, as seen in the middle panel of Fig.~\ref{fig:2}.   Figure~\ref{fig:2} shows that  the QPO  with period around 2.34~yr seems to dominate the time series after 2006. Following \cite{2022ApJ...926L..35O}, we also performed the LSP analysis only on the portion of the LC between 2006 and 2020.  We found a similar peak at frequency $\sim$0.43~yr$^{-1}$ (period: $\sim$~2.3~yr) in the LSP of this  temporal range. We again simulated $10^6$ LCs, using the same set of steps, estimating an average LSP power level. The lower panel of Fig.~\ref{fig:3} includes the LSP results of this portion of the LC.  We find that the peak is at 99.9976\% and 99.9984\% local significance levels with $\chi^2$ statistics and simulations, respectively. Following \cite{2005A&A...431..391V}, with $\sim$40 independent frequency trials, these significance levels respectively correspond to $\sim$99.90\% and $\sim$99.94\%, globally.
 
\section{REDFIT periodogram analysis}
\begin{figure}
    \centering
\includegraphics[width=1\linewidth]{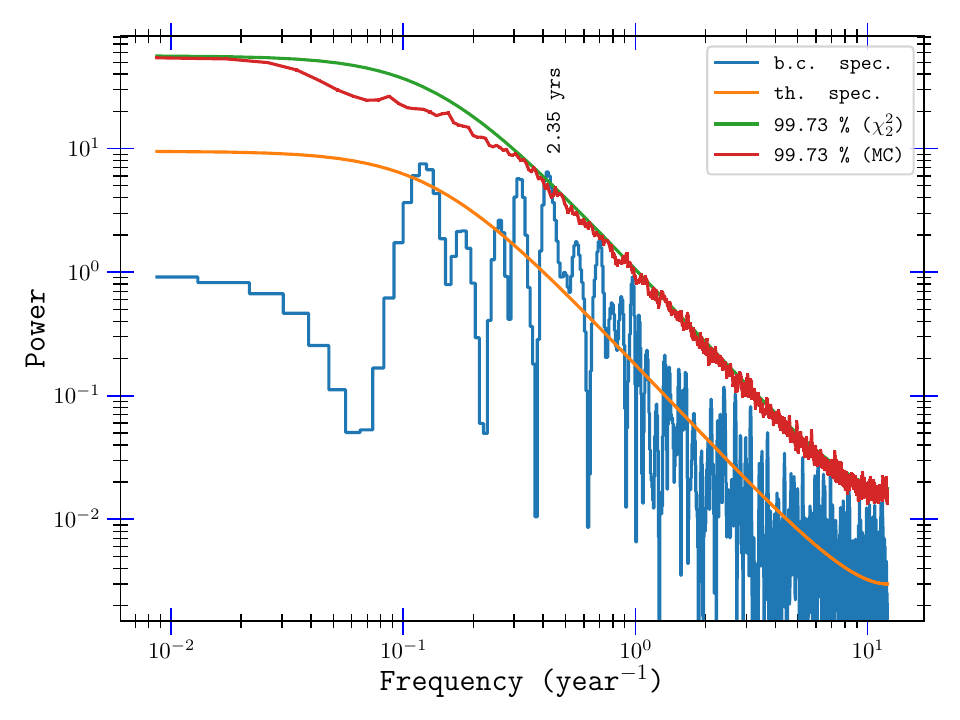}
        \caption{REDFIT plot of the segmented light curve of Ton 599 (1997 -- 2020): b.c. spec. and th. spec. denote the bias-corrected spectrum and the fitted AR1 model spectrum, respectively.}
    \label{fig:4}
\end{figure}    
\noindent
The power spectra of the blazars are often characterized by the red noise structure, because of the stochastic nature of their LCs, which can often be described by the simplest AutoRegressive process of first order (AR1, equivalent to a damped random walk or Ornstein-Uhlenbeck process). Sometimes higher-order processes have also been employed to fit more complex LC behaviours of astronomical objects; these include ARMA, ARIMA, ARFIMA, and CARMA \citep[see e.g.][and references therein]{2014ApJ...788...33K, 2018FrP.....6...80F, 2020ApJS..250....1T}. The AR1 model has often been utilized as a null hypothesis to understand whether the observed variations are of stochastic origin, and is given as
\begin{equation}
X(t_i) = \rho_iX(t_{i-1})+\epsilon(t_i)
\end{equation}
and
\begin{equation}
\rho_i = exp[-(t_i-t_{i-1})/\tau],
\end{equation}
where \(X(t_i)\) is the flux value at the timestamp \(t_i\), \(\rho~\text{and}~\tau\) are the two AR1 parameters, \(\epsilon\) is the white noise with zero mean and variance given as \(1-exp[-2(t_i-t_{i-1})/\tau]\).
The REDFIT software is a FORTRAN-based package allowing for efficient estimation of the AR1 parameters from unevenly sampled time series, avoiding any interpolation or spectrum bias arising from that sampling \citep[][]{2002CG.....28..421S}. This method is able to estimate red noise background power spectrum by fitting the AR1 model and can be efficiently used to test the genuineness of any frequency peaks against the estimated spectrum based on the \(\chi^2\) distribution for various significance levels. In addition, significance levels are calculated from the percentiles of the ensembles via Monte Carlo (MC) simulations.\\
\\
We employed the REDFIT analysis on Ton~599 time series data for the interval 1997 -- 2020 and present the results in Fig.~\ref{fig:4}, which includes the power spectra of the fitted AR1 model over that of the given LC, and 99.73\% significance levels with two methods. The plot indicates a strong presence of non-AR1 components in the LC which cross the 99.73\% significance levels at \(\sim\)0.43~yr\(^{-1}\) (\(\sim\)2.4~yr), which can be considered as a periodicity signature.

\section{Comparision of radio QPO with optical and $\gamma$-ray data}
\label{optical_gamma}
\subsection{Optical data and analysis}
\begin{figure}[h]
    \centering
    \includegraphics[width=1\linewidth, trim = 0 0 0 0.5cm, clip]{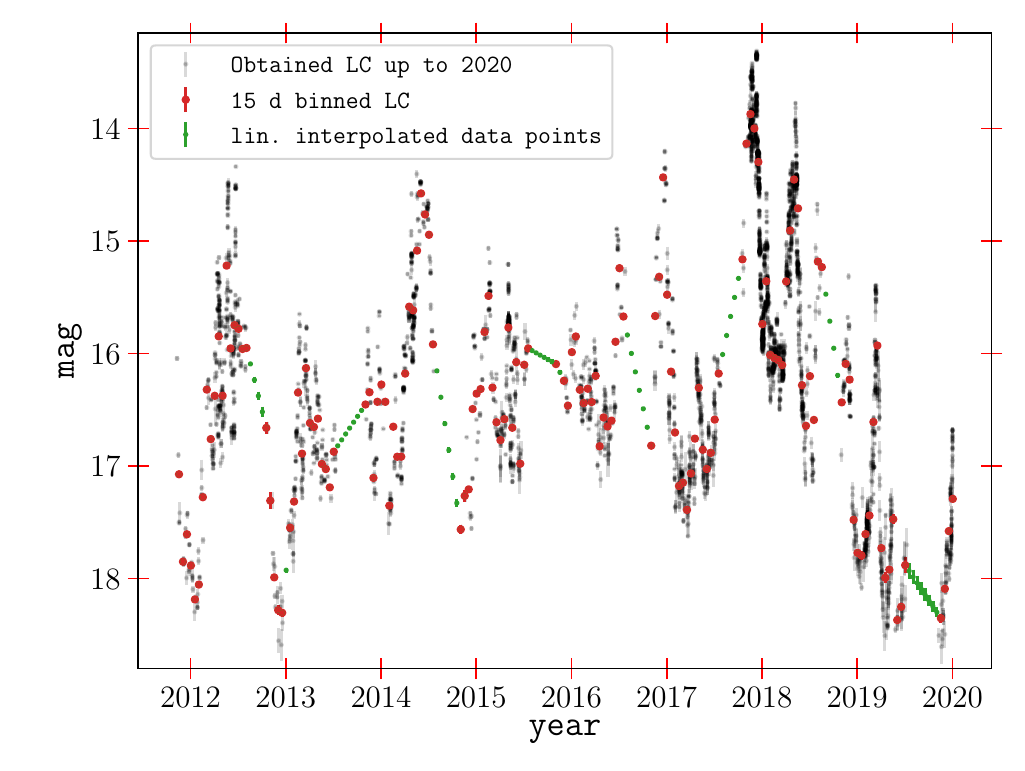}
    \includegraphics[width=1\linewidth, trim = 0 0 0 0.5cm, clip]{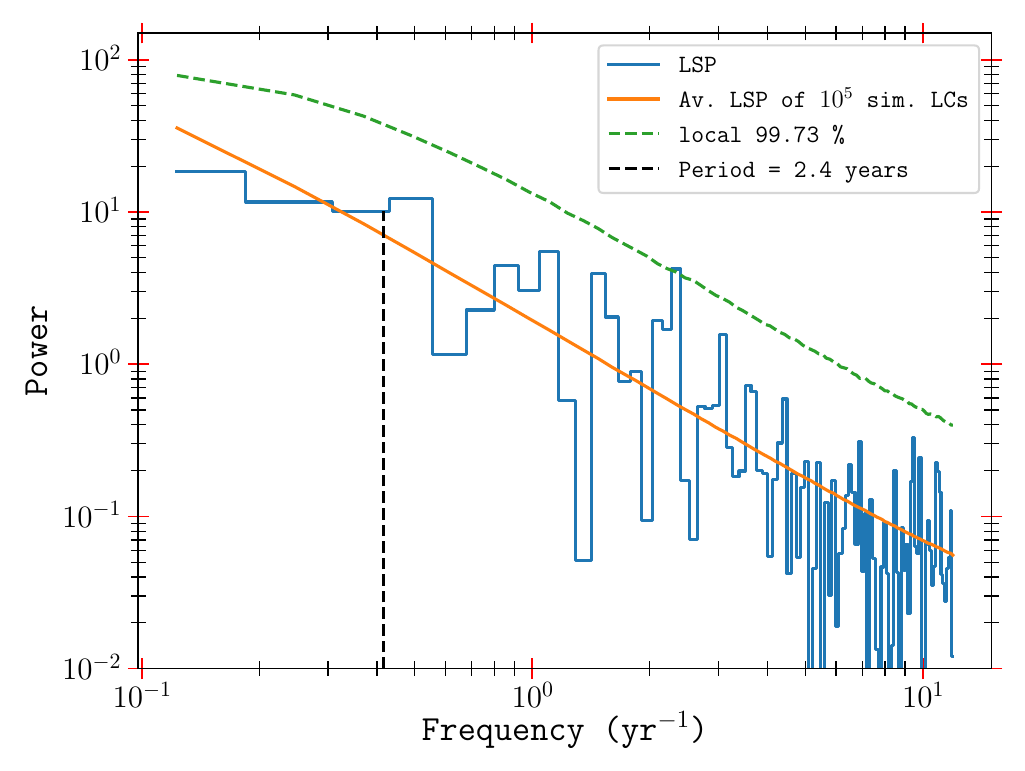}
    \caption{Upper panel: R-band light curve from \citet {2025A&A...703A.259V} binned into 15-day intervals, spanning from November 2011 to December 2020, and including points obtained with linear interpolation in gaps. Lower panel: LSP evaluation of the R-band binned and interpolated light curve.}
    \label{R_band_LC_LSP}
\end{figure}
\noindent
Recently a multi-band optical photometric variability study on the diverse timescales of the blazar Ton 599 was published by the Whole Earth Blazar Telescope (WEBT) consortium \citep{2025A&A...703A.259V}. The paper covers densely sampled B, V, R, and I optical band light curves of the source using data taken from 2011 to 2023. To compare any QPO behaviour with similar timescales in the optical regime with our radio observations, we utilized the densest sampled R-band light curve data from 2011 to 2020 for the present study.
The light curve data included lots of unevenness in sampling, along with a number of data gaps, which could lead to red noise leakages. Thus, a 15-day binning followed by linear interpolation was implemented to ameliorate these issues. The top panel in Fig.~\ref{R_band_LC_LSP} represents the light curves thus obtained. We then employed a LSP test and determined a 99.73\% significance level (via simulation method as done with the radio data) for any peculiar variations present in the periodogram. The PSD of the optical data was found to be a simple power law. We simulated $10^5$ light curves based on the flux distribution and PSD shape of the R-band light curve to get a local $3\sigma$ level. However, we found no obvious variations or LSP peak that could reach anywhere near the $3\sigma$ level in the vicinity of the QPO period of 2.4 years, in agreement with the analysis in \citet{2025A&A...703A.259V}. The lower panel of Fig.~\ref{R_band_LC_LSP} illustrates our evaluation, including the LSP of the R-band light curve, average LSP of $10^5$ simulated light curves and the 99.73 percentile significance level.
 
\subsection{Gamma-ray data and analysis}
\begin{figure}
    \centering
    \includegraphics[width=1\linewidth]{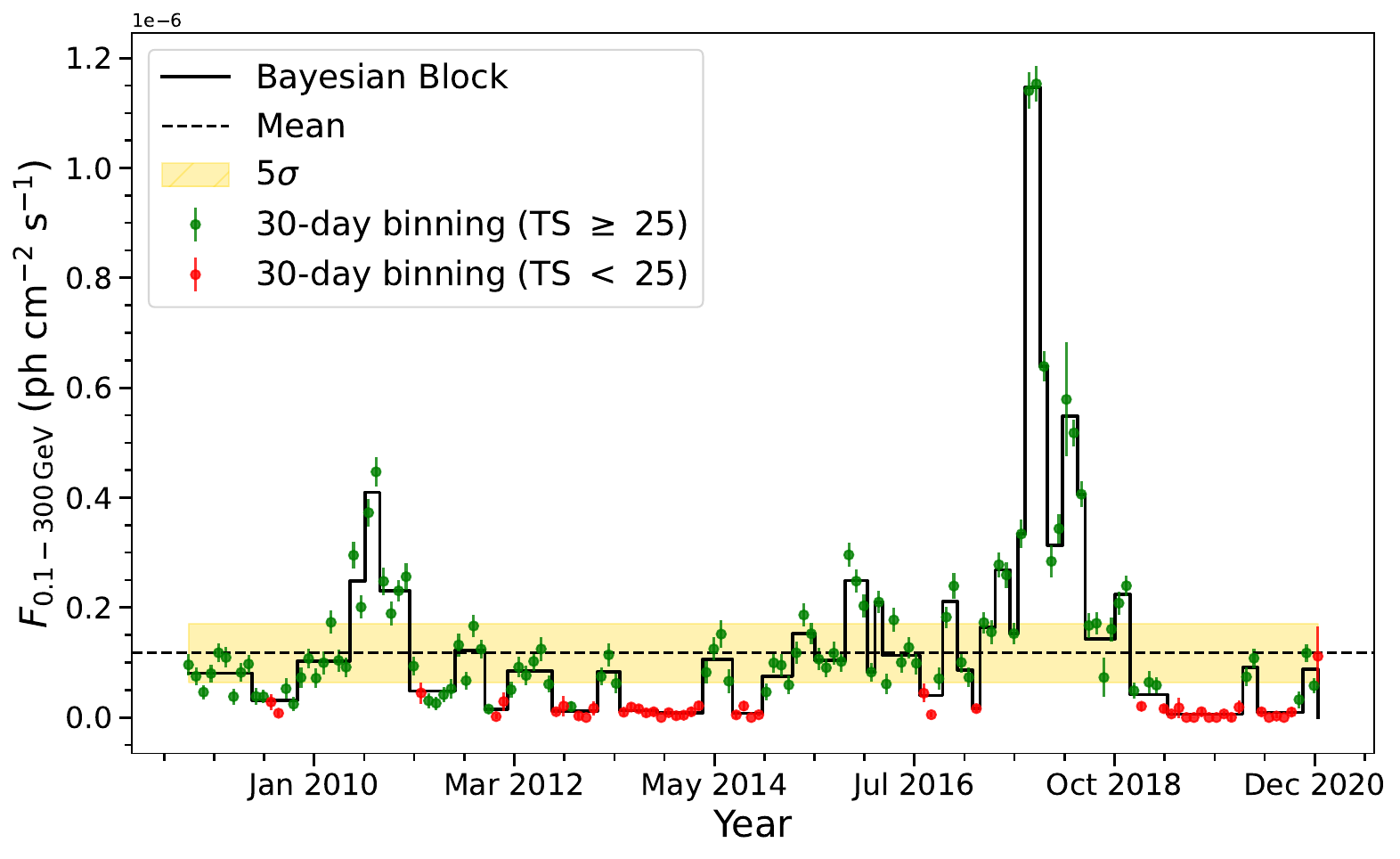}
    \includegraphics[width=1\linewidth]{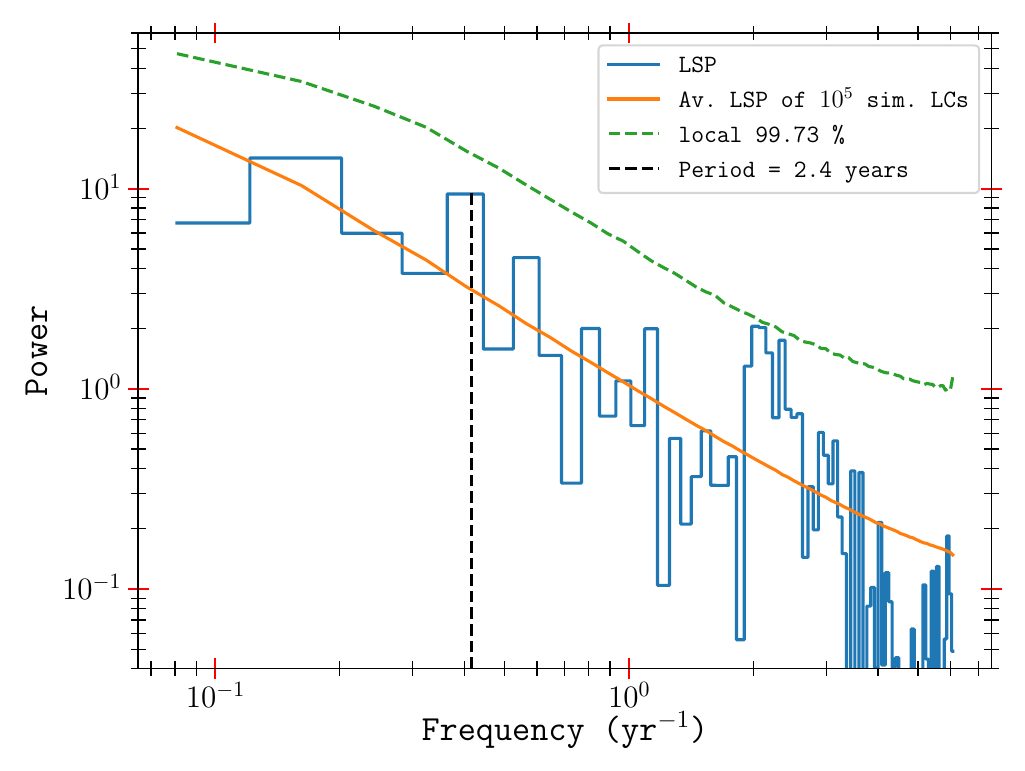}
    \caption{Upper panel: $\gamma$-ray light curve using a Bayesian Block method to detect enhanced $\gamma$-ray states with flares. In the label, TS denotes the test statistic, measuring how significantly a source is detected in a given time bin from the likelihood analysis. A value above 25 indicates a solid detection. Lower panel: LSP evaluation along with its significance estimation.}
    \label{gamma_LC_LSP}
\end{figure}
\noindent
Some details about the Fermi Gamma-ray telescope, the instruments on board, and its survey mode observation method are described in \citet{2025arXiv251223416M}. We retrieved the Pass 8 \textit{Fermi}-LAT $\gamma$-ray data for Ton 599 from the Fermi Science Support Centre (FSSC) data server \citep{fermi_lat_query}, covering almost 12.3 years (4 August 2008 to 31 December 2020). We extracted the counts within a circular region of interest with a radius of 20$^\circ$ centred on Ton 599, in the energy range from 100 MeV to 300 GeV. With all the standard procedures using Fermipy \citep[v1.0.1:][]{Fermipy_Version}, following \cite{Atwood_2009, bruel2018fermi, 2023arXiv230712546B, Fermi-LAT_galactic_diffuse_model, Bayesian_Block_2013}, an evenly sampled 30-day binned light curve, spanning 12.3 years, was obtained and has been presented in the upper panel of Fig.~\ref{gamma_LC_LSP}.\\ 
\\
We performed the $\gamma$-ray light curve analysis as for the optical R-band data, and evaluated its PSD to search for the presence any peculiar peak near the 2.4~yr radio periodicity. In addition, we also simulated $10^5$ light curves, based on $\gamma$-ray flux distribution and PSD shape, to get an ensemble of LSPs for significance level testing. The PSD of this light curve was also found to follow a simple power for this temporal interval. The LSP results are presented in the lower panel of Fig.~\ref{gamma_LC_LSP}. Similar to the negative result from the optical R-band for a LSP peak, we found no significant detection of any periodicity in vicinity of 2.4~yr in these Fermi-LAT data.
 \end{appendix}
\end{document}